\documentclass[twocolumn]{aastex7}

\usepackage[utf8]{inputenc}
\usepackage[english]{babel}
\usepackage{graphicx}

\newcommand{\pipe}{PIPE}
\newcommand{\piii}{PETRA\,III}

\begin{document}

\title{Double to tenfold $M$-shell photoionization of singly charged lanthanum ions}

\author[orcid=0009-0009-7009-3859, gname=Mirko, sname=Looshorn]{Mirko Looshorn}
\affiliation{I. Physikalisches Institut, Justus-Liebig Universt{\"a}t Gie{\ss}en, Heinrich-Buff-Ring 16, 35392 Giessen, Germany}
\affiliation{Helmholtz Forschungsakademie Hessen f{\"u}r FAIR (HFHF), GSI Helmholtzzentrum f{\"u}r Schwerionenforschung, Campus Gie{\ss}en, Heinrich-Buff-Ring 16, 35392 Giessen, Germany}
\email{mirko-dieter.looshorn@physik.uni-giessen.de}

\author[orcid=0009-0001-3405-5910, gname= B. Michel, sname=D\"ohring]{B. Michel D\"ohring}
\affiliation{I. Physikalisches Institut, Justus-Liebig Universt{\"a}t Gie{\ss}en, Heinrich-Buff-Ring 16,  35392 Giessen, Germany}
\affiliation{Helmholtz Forschungsakademie Hessen f{\"u}r FAIR (HFHF), GSI Helmholtzzentrum f{\"u}r Schwerionenforschung, Campus Gie{\ss}en, Heinrich-Buff-Ring 16, 35392 Giessen, Germany}
\email{michel.doehring@exp1.physik.uni-giessen.de}

\author[orcid=0000-0003-0166-2666, gname=Pierre-Michel, sname=Hillenbrand]{Pierre-Michel Hillenbrand}
\affiliation{I. Physikalisches Institut, Justus-Liebig Universt{\"a}t Gie{\ss}en, Heinrich-Buff-Ring 16, 35392 Giessen, Germany}
\affiliation{GSI Helmholtzzentrum f\"u{}r Schwerionenforschung, Planckstra{\ss}e 1, 64291 Darmstadt, Germany}
\email{P.M.Hillenbrand@gsi.de}

\author[orcid=0000-0002-1228-5029, gname=Michael, sname=Martins]{Michael Martins}
\affiliation{Institut f\"{u}r Experimentalphysik, Universit\"{a}t Hamburg, Luruper Chaussee 149, 22761 Hamburg, Germany}
\email{michael.martins@uni-hamburg.de}

\author[orcid=0000-0002-0030-6929, gname=Alfred, sname=M\"uller]{Alfred M\"{u}ller}
\affiliation{I. Physikalisches Institut, Justus-Liebig Universt{\"a}t Gie{\ss}en, Heinrich-Buff-Ring 16,  35392 Giessen, Germany}
\email{Alfred.Mueller@iamp.physik.uni-giessen.de}

\author[orcid=0000-0001-8106-7124,gname=Simon, sname=Reinwardt]{Simon Reinwardt}
\affiliation{Institut f\"{u}r Experimentalphysik, Universit\"{a}t Hamburg, Luruper Chaussee 149, 22761 Hamburg, Germany}
\email{simon.reinwardt@desy.desimon.reinwardt@desy.de}

\author[orcid=0009-0009-7545-2101,gname=J\"orn, sname=Seltmann]{J\"orn Seltmann}
\affiliation{DESY Photon Science, Deutsches Elektronen-Synchrotron, Notkestraße 85, 22607 Hamburg, Germany}
\email{joern.seltmann@desy.de}

\author[orcid=0000-0002-0891-9180, gname=Florian, sname=Trinter]{Florian Trinter}
\affiliation{Molecular Physics, Fritz-Haber-Institut der Max-Planck-Gesellschaft, Faradayweg 4-6, 14195 Berlin, Germany}
\email{florian.trinter@desy.de}

\author[orcid=0000-0002-6305-3762, gname=Shuxing, sname=Wang]{Shuxing Wang}
\affiliation{I. Physikalisches Institut, Justus-Liebig Universt{\"a}t Gie{\ss}en, Heinrich-Buff-Ring 16,  35392 Giessen, Germany}
\affiliation{Helmholtz Forschungsakademie Hessen f{\"u}r FAIR (HFHF), GSI Helmholtzzentrum f{\"u}r Schwerionenforschung, Campus Gie{\ss}en, Heinrich-Buff-Ring 16, 35392 Giessen, Germany}
\email{shuxing.wang@phyisk.uni-giessen.de}

\author[orcid=0000-0003-2441-4075, gname=Aloka Kumar, sname=Sahoo]{Aloka Kumar Sahoo}
\affiliation{GSI Helmholtzzentrum f\"u{}r Schwerionenforschung, Planckstra{\ss}e 1, 64291 Darmstadt, Germany}
\affiliation{Helmholtz-Institut Jena, Fr{\"o}belstieg 3, 07743 Jena, Germany}
\affiliation{Theoretisch-Physikalisches Institut, Friedrich-Schiller-Universit\"{a}t Jena, Max-Wien-Platz 1, 07743 Jena, Germany}
\email{aloka_s@ph.iitr.ac.in}

\author[orcid=0000-0003-3101-2824, gname=Stephan, sname=Fritzsche]{Stephan Fritzsche}
\affiliation{GSI Helmholtzzentrum f\"u{}r Schwerionenforschung, Planckstra{\ss}e 1, 64291 Darmstadt, Germany}
\affiliation{Helmholtz-Institut Jena, Fr{\"o}belstieg 3, 07743 Jena, Germany}
\affiliation{Theoretisch-Physikalisches Institut, Friedrich-Schiller-Universit\"{a}t Jena, Max-Wien-Platz 1, 07743 Jena, Germany}
\email{s.fritzsche@gsi.de}

\author[orcid=0000-0002-6166-7138, gname=Stefan, sname=Schippers]{Stefan Schippers}
\affiliation{I. Physikalisches Institut, Justus-Liebig Universt{\"a}t Gie{\ss}en, Heinrich-Buff-Ring 16,  35392 Giessen, Germany}
\affiliation{Helmholtz Forschungsakademie Hessen f{\"u}r FAIR (HFHF), GSI Helmholtzzentrum f{\"u}r Schwerionenforschung, Campus Gie{\ss}en, Heinrich-Buff-Ring 16, 35392 Giessen, Germany}
\email{stefan.schippers@uni-giessen.de}

\correspondingauthor{Stefan Schippers}
\email{stefan.schippers@uni-giessen.de}

\begin{abstract}
Using the photon-ion merged-beams technique at the PETRA\,III synchrotron light source, we have measured cross sections for double and up to tenfold photoionization of La$^{+}$ ions by a single photon in the energy range 820--1400~eV, where resonances and thresholds occur that are associated with the excitation or ionization of one $M$-shell electron. These cross sections represent experimental benchmark data for the further development of quantum theoretical methods, which will have to provide the bulk of the atomic data required for the modeling of nonequilibrium plasmas such as kilonovae. In the present work, we have upgraded the Jena Atomic Calculator (JAC) and pushed the state-of-the-art of quantum calculations for heavy many-electron systems to new limits. In particular, we  have performed large-scale calculations of the La$^+$ photoabsorption cross section and of the deexcitation cascades, which set in after the initial creation of a $3d$ hole. Our theoretical results largely agree with our experimental findings. However, our theoretical product-ion charge state distributions are somewhat narrower than the experimental ones, which is most probably due to the simplifications necessary to keep the cascade calculations tractable.
\end{abstract}



\section{Introduction}

\begin{deluxetable}{lrccll}
\tablewidth{0pt}
 \tablecaption{\label{tab:PI}List of  experimental cross sections for photoionization of positive atomic ions with nuclear charge $Z>54$, i.e.,   heavier than xenon, that are available from the literature. $h\nu$ denotes the photon energy.}
\tablehead{
\colhead{Element} & \colhead{$Z$} & \colhead{charge state} & \colhead{$h\nu$ range (eV)} & \colhead{Degree of ionization} & \colhead{Reference}
}
\startdata
  Cs &55 &  1      & \phantom{1}90 -- \phantom{1}160 & single \& double  &\cite{Kjeldsen2002b}\\
  Cs &55 & 1       & \phantom{1}80 -- \phantom{1}140 & single \& double  &\cite{Koizumi2009}\\
  Ba &56 & 1       & \phantom{1}16 -- \phantom{11}30 & single  &\cite{Lyon1986} \\
  Ba &56 & 1       & \phantom{1}70 -- \phantom{1}190 & single \& double  &\cite{Koizumi1995} \\
  Ba &56 & 1       & \phantom{1}90 -- \phantom{1}160 & single, double \& triple &\cite{Kjeldsen2002b} \\
  Ba &56 & 2       & \phantom{1}90 -- \phantom{1}160 & single \& double  &\cite{Kjeldsen2002b} \\
  Ba &56 & 2 -- 3  &           100 -- \phantom{1}150 & single \& double  &\cite{Bizau2001} \\
  Ba &56 & 4 -- 5  &           100 -- \phantom{1}150 & single  &\cite{Bizau2001} \\
  Ce &58 & 1 -- 2  &           105 -- \phantom{1}170 & single, double \& triple   &\cite{Habibi2009} \\
  Ce &58 & 3  &                105 -- \phantom{1}170 & single \& double  &\cite{Habibi2009} \\
  Ce &58 & 4 -- 9  &           105 -- \phantom{1}170 & single  &\cite{Habibi2009} \\
  Sm &62 & 2       &           100 -- \phantom{1}170 & single \& double  &\cite{Champeaux2004a} \\
  Sm &62 & 3       & \phantom{1}80 -- \phantom{1}200 & single \& double  &\cite{Bizau2012} \\
  Eu &63 & 1       &           110 -- \phantom{1}160 & single \& double  &\cite{Kojima1998}\\
  W & 74 &1        & \phantom{1}16 -- \phantom{1}245 & single  &\cite{Mueller2015c} \\
  W & 74 &2 -- 3   & \phantom{1}20 -- \phantom{11}90 & single  &\cite{McLaughlin2016a} \\
  W & 74 &4        & \phantom{1}40 -- \phantom{1}105 & single  &\cite{Mueller2017a} \\
  W & 74 &5        & \phantom{1}20 -- \phantom{1}160 & single   &\cite{Mueller2019a} \\
\enddata
\end{deluxetable}

In 2017, the LIGO/Virgo collaboration detected the first gravitational-wave signal from the GW170817 merger of a neutron-star binary \citep{Abbott2017}. Less than two seconds later, a  short gamma-ray burst was detected, followed by a several-days-lasting optical “afterglow” (AT2017gfo) powered by the radioactive decay of the neutron-rich material ejected in the merger, i.e., a kilonova \citep{Metzger2020}. The kilonova light-curves and spectra hint to large abundances of heavy elements which must have been produced in the energetic neutron-star merger event \citep{Kasen2017,Watson2019,Holmbeck2023}. In order to be able to quantify the elemental abundances, atomic data are required for the basic atomic processes that occur in kilonovae. This need for data has triggered a rapidly growing number of mainly theoretical studies on the atomic properties of heavy elements (lanthanides, actinides) \citep[e.g.,][]{Radziute2020,Domoto2021,CarvajalGallego2022,Banerjee2022,Domoto2022, BenNasr2023,Bondarev2023,BenNasr2024,Gaigalas2024,Deprince2025}.

So far, local thermodynamic equilibrium (LTE) conditions have mostly been assumed in the astrophysical modeling of kilonovae  \citep{Metzger2010,Kasen2017,Watson2019,Tanaka2013}, which is certainly an over\-simpli\-fica\-tion given the highly dynamic and transient nature of the phenomenon. Only very recently, the impact of non-LTE effects on kilonovae has been estimated \citep{Hotokezaka2021,Pognan2023} highlighting the need of accurate atomic cross sections and rate coefficients. Such quantities cannot, in general, be calculated with sufficient precision for the heavy many-electron ions of interest. The currently available atomic data for heavy elements stem mostly from theoretical calculations with limited (usually not quantified) accuracy. Experiments which challenge such theoretical work are coming up only slowly \citep[see, e.g.,][]{Dowd2025}. 

The present study on the inner-shell photoionization of singly charged lanthanum ions belongs to our greater effort to meet these atomic-data needs by providing benchmark cross sections for atomic collision processes such as electron-impact excitation and ionization, electron-ion recombination, and photoionization of low-charged lanthanide (and heavier) ions \citep{Doehring2025}. 
The process of  $(q\!-\!1)$-fold photoionization of La$^+$ ions can be written as
\begin{equation}\label{eq:reaction}
h\nu + \mathrm{La}^+ \to \mathrm{La}^{q+} + (q-1)e^-
\end{equation}
where $h\nu$ denotes the photon energy and $q$ is the charge state of the product ion.  In addition to the respective experimental cross sections $\sigma_q$ for $3\leq q \leq 11$, the present study also provides theoretical absorption cross sections and a theoretical treatment of the deexcitation cascades that set in after the initial creation of an inner-shell hole. 

The challenge for atomic theory lies in the manifest many-particle character of the atomic systems of interest, i.e., low to moderately charged atomic ions of heavy elements with typically 50 or more electrons.  If atomic data for heavy many-electron systems exist (mostly from theoretical calculations), their accuracy is often questionable. This concerns basic atomic quantities such as levels energies, transition rates \citep{Quinet2020}, and cross sections of atomic collision processes. Critical comparisons, such as the present one,   between experiment and theory are required for guiding the further development of the theoretical methods.  For photoionization of heavy ions, the experimental data base is rather poor. Table~\ref{tab:PI} provides a comprehensive list of experimental photoionization studies with atomic ions heavier than xenon, that have been performed so far. Their number is modest. 

In the following Section \ref{sec:exp}, we describe the experimental method that was applied for measuring the cross sections for multiple photoionization of La$^+$. In Section \ref{sec:res}, we present our experimental cross sections and compare these with the results of our atomic quantum calculations. Finally, Section \ref{sec:sum} provides a summary and conclusions.     

\section{Experimental Method}\label{sec:exp}

The experiment was conducted at the soft x-ray photon beamline P04 of the \piii\ synchrotron radiation facility operated by DESY in Hamburg, Germany. We used the permanently installed photon-ion end station at \piii\ (\pipe) where the La$^+$ ion beam was merged with the counter propagating P04 photon beam. The \pipe\ setup and the experimental procedures for measuring  photoionization cross sections have been described in detail by \cite{Schippers2014} and \citet{Mueller2017}.  In the past decade, the \pipe\ setup produced photoionization cross sections for a number of ions of astrophysical interest \citep{Schippers2020c}. The latest related work comprises $L$-shell photoionization of  Fe$^{2+}$ \citep{Schippers2021} and Ar$^+$ \citep{Mueller2023} as well as $K$-shell photoionization of Si$^{+}$, Si$^{2+}$, Si$^{3+}$ \citep{Schippers2022}, C$^{2+}$ \citep{Mueller2023},  B$^{3+}$ \citep{Mueller2024,Mueller2025}, and B$^{2+}$ \citep{Mueller2025a}. 

In the present experiment, La$^+$ ions were produced from solid lanthanum metal in a Penning discharge ion source operated in sputter mode \citep{Baumann1981}. The ion source was on a positive electrical potential $U_\mathrm{acc} = 6$~kV and the ions were extracted towards the electrically grounded ion beamline. The extracted ion beam was passed through a dipole magnet for selecting the desired $^{139}$La$^+$ species according to its mass-to-charge ratio $m/e=138.906$~u/$e$ \citep{Wang2021c} with $e$ denoting the elementary charge. Employing an electrostatic deflector, the La$^+$ ion beam was subsequently bent onto the axis of the photon beam and collimated by two sets of four-jaw slits located at the entrance and the exit of the interaction region. These slits were closed such that the photon flux was just not intercepted. The resulting beam sizes were about 2$\times$2~mm$^2$, and the beam overlap was practically independent of the photon energy. The collimated ion current was in the range of 1--2~nA. 

Using a monochromator grating with 1200 lines per millimeter and a monochromator exit-slit width of 1000~$\mu$m, the photon flux was $\sim$ $4\times10^{13}$~s$^{-1}$ across the experimental photon-energy range. The length of the overlap between the two beams was approximately 1.7~m.  La$^{q+}$ product ions with $3\leq q \leq 11$ resulting from multiple photoionization of La$^+$ (cf.~equation \ref{eq:reaction}) were magnetically separated from the primary beam and directed onto a single-particle detector, the efficiency of which is close to 100\% independent of $q$ \citep{Rinn1982}. 

Product-ion yields were measured as functions of photon energy separately for each of the  product charge states $q$. The yields were obtained by normalizing the background-subtracted La$^{q+}$ count rates to the La$^+$ parent-ion current, which was measured with a Faraday cup, and on the photon flux, which was recorded  with a calibrated photodiode. The background count rates were measured separately for each product-ion channel with the photon beam being blocked from entering the PIPE setup. 

In principle, the product-ion yields can be put on an absolute total photoabsorption scale by additionally accounting for the geometric overlap of the beams \citep{Schippers2014}. However, a measurement of this quantity is rather time consuming and was omitted in view of the limited beamtime available. Instead, we scaled the experimental product-ion yields to a  theoretical total photoabsorption cross section using an energy-independent scaling factor as explained below. Correspondingly, the uncertainty of our absolute cross section scale is the $\pm$20\% uncertainty of this theoretical cross section.

\begin{figure}[b]
	\vspace*{14pt}\centering\includegraphics[width=0.85\columnwidth]{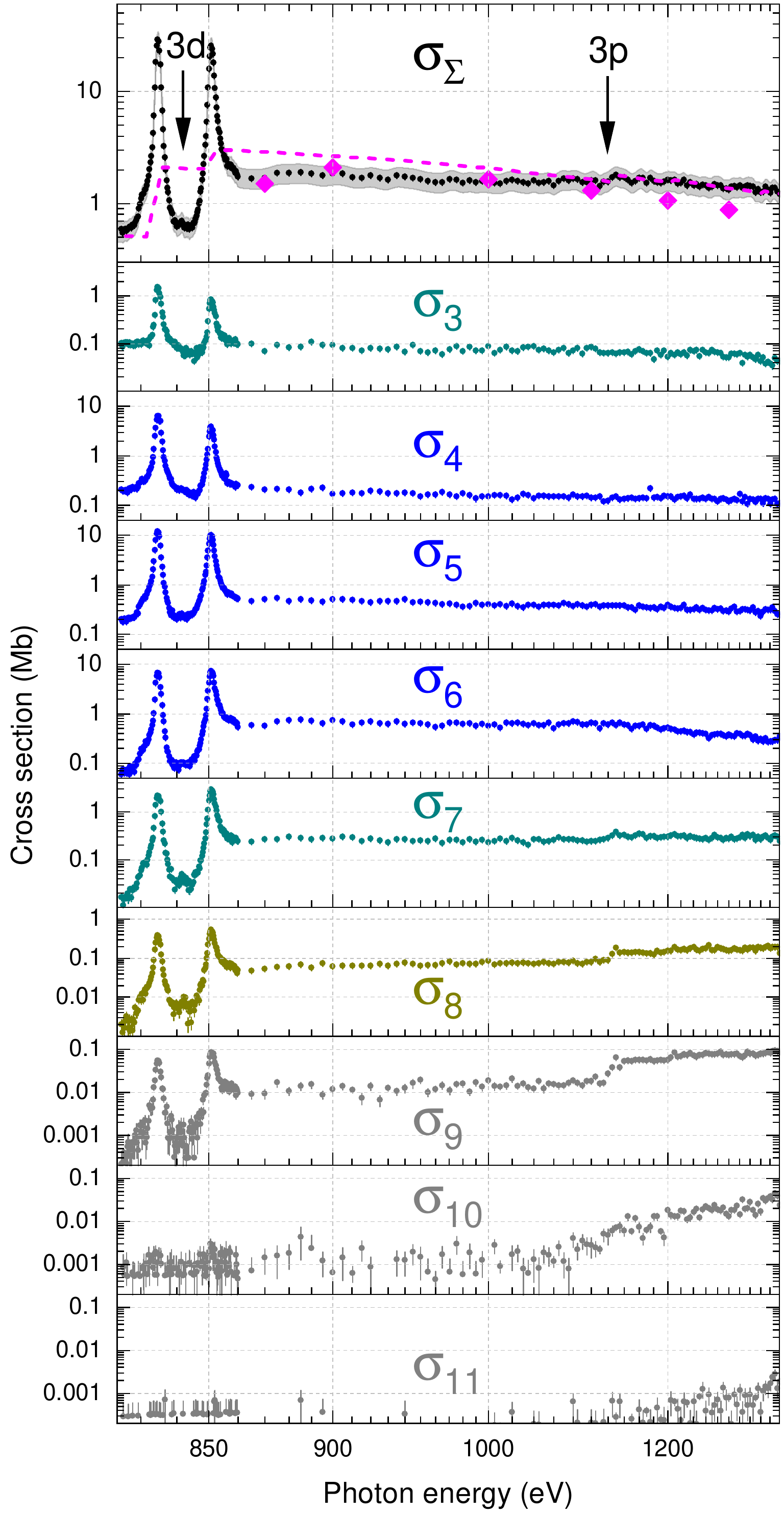}
	\caption{\label{fig:all} Experimental cross sections (symbols with statistical error bars) $\sigma_q$ for net ($q$-1)-fold photoionization of La$^+$ ions in the photon-energy range 824--1400~eV. The photon-energy ($h\nu$) axis is compressed toward higher energies using $x = \log(h\nu/\mathrm{eV} - 750)$ for the abscissa in order to provide a better view of the resonance structures at lower energies. The various panels have different cross section scales. Cross sections of the same color are plotted on the same scale.  The top panel shows the sum cross section $\sigma_\Sigma$ (symbols), which was scaled to the theoretical absorption cross section for neutral La atoms of \citet{Chantler2005} (dashed magenta line) to put the experimental cross sections on an absolute scale. The gray shaded band marks the $\pm$20\% systematic uncertainty of the thus calibrated experimental cross section. The magenta diamonds represent the present theoretical cross sections for direct photoionization. The vertical arrows mark the approximate locations of the $3d$ and $3p$ ionization thresholds. \textbf{The experimental data is available on Zenodo under an open-source 
Creative Commons Attribution license: \dataset[doi:10.5281/zenodo.17977213]{https://doi.org/10.5281/zenodo.17977213}.}}%
\end{figure}

The photon-energy scale was calibrated by measuring the carbon $1s\to\pi^*$ resonance in neutral CO  at 287.40(2)~eV \citep{Sodhi1984a} and the $1s\to3p$ resonance in neutral Ne at 867.3(2)~eV \citep{Mueller2017} and using a linear calibration function for the entire experimental energy range. The linearity of the photon-energy scale follows from the linearity of the rotations of the monochromator pre-mirror and the monochromator grating, which are monitored online by precision encoders \citep{Viefhaus2013}. Residual nonlinearities of the photon-energy scale increase with photon energy from  $\pm 10$~meV at 290~eV up to about 100~meV at 1400~eV  as measured by photoelectron spectroscopy similar to the method reported by \citet{Togawa2024}. These nonlinearities are lower than the uncertainties of the calibration lines. Finally, the Doppler shift due to the motion of the ions was taken into account  by multiplying the calibrated energies by the factor$\sqrt{(1+\beta)/(1-\beta)}\approx1.00305$, where $\beta=\sqrt{2eU_\mathrm{uacc}/mc^2} \approx 3.045\times10^{-4}$ denotes the ion velocity in units of the vacuum speed of light $c$.
The uncertainty of the calibrated and Doppler-corrected photon-energy scale amounts to $\pm0.2$~eV below 870~eV and up to $\pm$0.4~eV at 1400~eV.

The La$^+$ primary ion beam consisted of an unknown mixture of the [Xe]$\,5d^2\;^3F_2$ ground level and excited metastable levels of the even-parity $5d^2$, $5d\,6s$, and $6s^2$ configurations. Using the Jena Atomic Calculator \citep[JAC,][]{Fritzsche2019} we calculated that the lifetimes of the pertaining fine-structure levels range from a few milliseconds to about 200 seconds. These times are much longer than the ions' $\sim$100-$\mu$s flight time from the ion source to the photon-ion interaction region.  Therefore, we must assume that all 12 fine-structure levels of the lowest-lying $5d^2$, $5d\,6s$, and $6s^2$ configurations with excitation energies of up to $\sim$1.25~eV \citep{Kramida2024} were present in our La$^+$ primary ion beam.         

\section{Results}\label{sec:res}

Figure~\ref{fig:all} shows the present experimental cross sections $\sigma_q$ for net ($q$-1)-fold photoionization of La$^+$ ions resulting in the production of La$^{q+}$ ions with $3\leq q \leq 11$. The investigated photon-energy range of about 820--1400~eV comprises the thresholds for the direct ionization of a $3d$ or a  $3p$ electron. The creation of such an inner-shell hole leads to deexcitation cascades of radiative and autoionizing transitions, which eventually produce the observed wide distributions of product-ion charge states, which vary with the photon energy. 

\subsection{Absorption cross section}\label{sec:abs}

The cross sections were put on an absolute scale by normalizing the experimental sum cross section 
\begin{equation}\label{eq:sigsum}
\sigma_\Sigma = \sum_{q=3}^{11}\sigma_q
\end{equation}
at energies above 1130~eV to the theoretical absorption cross section for neutral lanthanum as provided by NIST \citep[upper panel of Figure~\ref{fig:all}]{Chantler2005} and by multiplying all individual cross sections $\sigma_q$ by the thus found (photon-energy-independent) cali\-bration factor. The adequateness of such an approach has been discussed by \citet{Schippers2022}. At energies well above subshell ionization thresholds, the cross sections for photoabsorption of inner shells are not very sensitive to the number of valence electrons. Therefore, the photoabsorption cross section of neutral lanthanum can be safely assumed to serve as a useful proxy for the photoabsorption for La$^+$ at energies well above the $3p$ threshold and well below the $3d$ thresholds. In principle, one has also to be aware of the fact, that the photon-beam size and, thus, the geometric overlap between ion beam and photon beam depends on the photon energy. In our normalization procedure we took this into account by adding a term linear in photon energy to the calibration factor. However, this term turned out to be negligible concerning the quoted systematic uncertainty of up to $\pm$20\%  of the NIST cross section above the $3p$ threshold at about 1130~eV \citep{Chantler2000b}.     

Next to the experimental sum cross section and the recommended absorption cross section for neutral lanthanum, the upper panel of Figure~\ref{fig:all} also shows  our present theoretical absorption cross sections (diamonds), which have been obtained by using the JAC code \citep{Fritzsche2019,Fritzsche2025b} and by considering only nonresonant direct single ionization of the $3d$, $4s$, $4p$, or $4d$ subshells. At photon energies from 900~eV up to below the $3p$ ionization threshold at about 1100~eV, our theoretical result agrees with the recommended one to within 25\%, with our JAC result being closer to the experiment than the recommended cross section by NIST \citep{Chantler2005}. At 870~eV our calculated result is on the experimental curve, while the NIST cross section is about twice as large. This discrepancy may be partly due to the fact that the NIST cross section is for neutral lanthanum and that the difference in the valence shell as compared to singly ionized lanthanum is expected to manifest itself strongest in the immediate vicinity of inner-shell ionization thresholds, where in addition the uncertainty of the NIST data is quoted to be the largest \citep[up to $\pm$40\%,][]{Chantler2000b}. Above the $3p$ ionization threshold our theoretical results underestimate the absorption cross section. Most probably, this is due to the neglect of $3p$ ionization in our calculations, which would have added a considerable degree of complexity to the computations.

\begin{figure}
		\centering\includegraphics[width=\columnwidth]{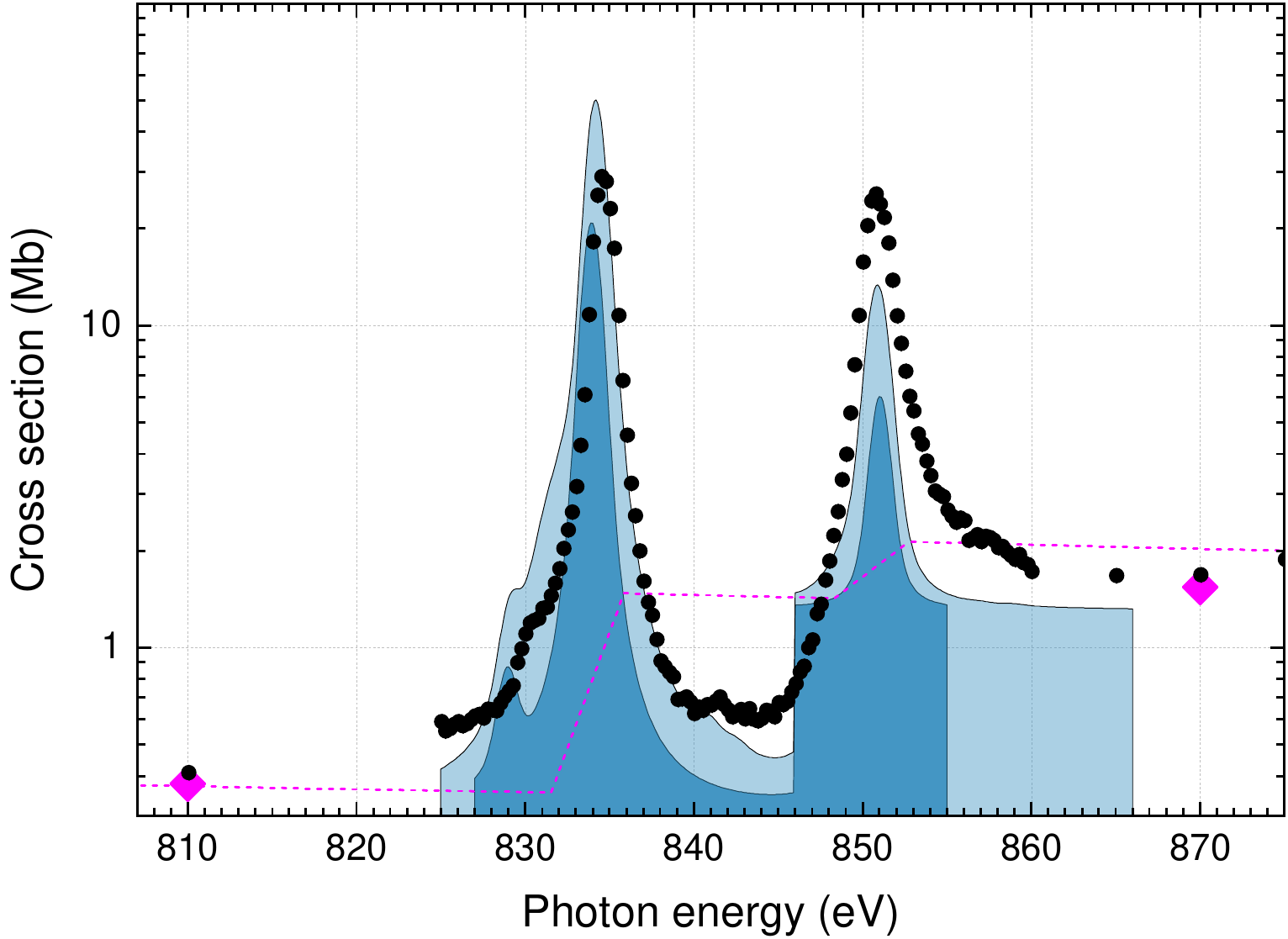}
	\caption{\label{fig:exptheo} Experimental sum cross section (black circles) and theoretical absorption cross section near the $3d$ ionization threshold. The dashed magenta line is the cross section by \citet{Chantler2005} multiplied by a factor of 0.7. The magenta diamonds represent the present theoretical cross section for direct ionization. The dark blue shaded curve is the calculated resonant cross section due to $3d\to 4f$ excitation for a [Xe]$\,6s^2$ initial configuration. The  light blue curve represents the resonant cross section accounting for configuration interaction between [Xe]$\,5d^2$ and [Xe]$\,4f^2$ initial configurations (shifted by -1~eV). Each cross section for resonant absorption consists of several resonances which were individually represented by Voigt profiles with 0.6-eV Lorentzian width and 1.0-eV Gaussian full width at half maximum. In order to (coarsely) account for direct ionization, constant offsets of 0.3~Mb and 1.3~Mb were added to the resonant cross sections below and above 846~eV, respectively.}        
\end{figure}

The prominent resonance features at $\sim$835~eV and $\sim$851~eV are associated with $3d_{5/2}\to4f$ and $3d_{3/2}\to4f$ excitations, respectively. This is very similar to  $3d$ photoionization of Xe$^+$ \citep{Schippers2014}, but the resonance features are much sharper for La$^+$ than for Xe$^+$. We attribute this to the higher nuclear charge of lanthanum, which leads to a more localized $4f$ subshell in La$^+$ as compared to Xe$^+$. The progression of the contributions of $3d\to nf$ resonances to the photoionization cross sections with increasing ion charge has already been studied in quite some detail for a range of multiply charged xenon ions \citep{Schippers2015a}.

The $3d\to4f$ assignment of the experimentally observed resonance features is supported by our present JAC calculations for resonant photoabsorption, which are displayed in Figure~\ref{fig:exptheo} together with our experimental sum cross section. In the calculations we have gradually increased the level of complexity. The simplest approach, where we have assumed a closed-subshell [Xe]$\,6s^2$ initial and a $3d^{-1}\,4f\,6s^2$ resonance configuration, already reproduces the observed fine-structure splitting. For the comparison with the experiment the individual resonances from all our calculations have been represented by Voigt profiles with a 1-eV Gaussian full width at half maximum, which accounts for the experimental photon-energy spread. For the natural Lorentzian line width we have used an average value of 0.6~eV throughout. An \textit{ab initio} calculation of the line widths was not attempted since this would have required a large-scale treatment of the many decay channels that are available for the $3d$ core-hole levels.  Despite these simplifications, the agreement between the theoretical and the experimental resonance cross section is encouraging.

In a more complex approach we accounted for the configuration mixing between the [Xe]$\,5d^2$ and the  [Xe]$\,4f^2$ initial configurations. This resulted in remarkably good agreement between experiment and theory if all resonance energies are shifted by $-1$~eV. This shift is larger than the uncertainty of the experimental photon-energy scale but within the uncertainty of the present theoretical approach. 

The theoretical uncertainty behind the rather uniform $-1$~eV energy shift is consistent with the expected accuracy of present-day atomic-structure methods for these medium-heavy and strongly correlated La$^+$ ions. This shift compensates for well-known limitations in the computation of inner-shell hole states and their subsequent autoionization, a situations that is further complicated by the open $4f$ shell. Open $f$-shell ions are known for strong configuration interaction that may significantly perturb all intermediate resonances following the photoabsorptions as well as the further electron emission towards the final ground state of the ions. Therefore, the excitation energies of inner-shell electrons typically deviate by 1--2~eV depending due to an  imbalance between different correlation contributions.

For the (scaled) photoabsorption cross sections, the overall uncertainty is necessarily large and amounts to about 30\%, with sometimes even larger deviations. No systematic improvement of wave functions is possible for such ions with complex shell structure. We therefore estimate  conservatively  the uncertainty of the theoretical cross sections to be up to 50\%, though with variations depending on the energies of the incoming photons.

We conclude this section by noting that the JAC code is obviously capable of providing reasonably  accurate cross sections for the photoabsorption of many-electron heavy atoms such as, presently, La$^+$, despite the fact, that the complex nature of the many-electron systems under study with multiple open shells requires one to strictly limit the sets of configurations to be considered to keep the calculations tractable.

\subsection{Ionization cross sections}

The various measured multiple-ionization cross sections displayed in Figure~\ref{fig:all} span more than four orders of magnitude ranging from 1~kb to more than 10~Mb. The largest cross sections are $\sigma_4$, $\sigma_5$, and $\sigma_6$ for threefold to fivefold ionization. The cross sections $\sigma_q$ for $3\leq q \leq9$ exhibit the above-discussed resonance features at $\sim$835eV and $\sim$851~eV and an almost flat continuum at higher energies. Both the resonances and the continuum are very weak or missing in $\sigma_{10}$ and $\sigma_{11}$ which significantly rise only at energies above about 1100~eV.

\begin{figure}
		\centering\includegraphics[width=\columnwidth]{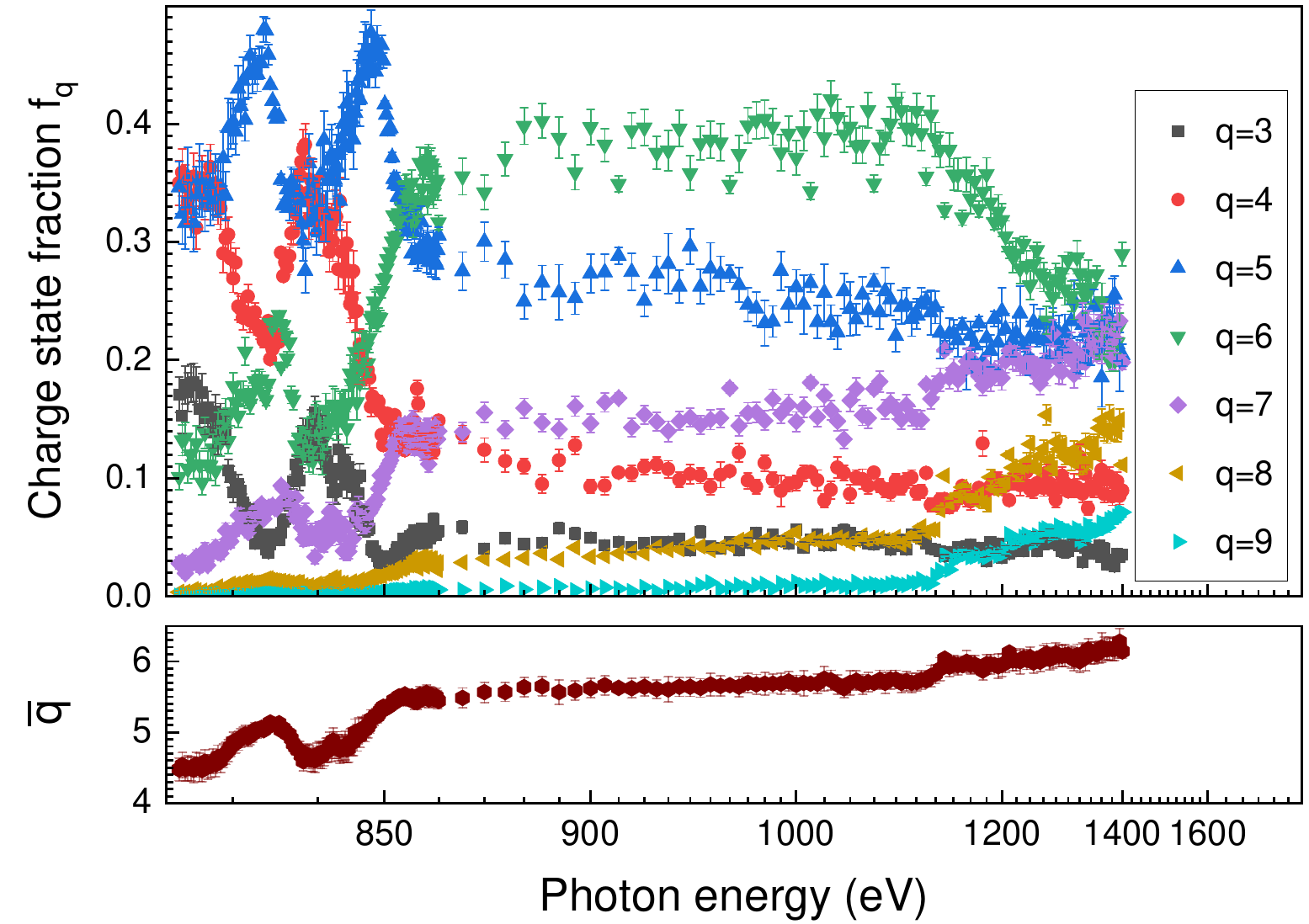}
	\caption{\label{fig:frac}Experimental charge state fractions $f_q$ (Equation~\ref{eq:fq}, upper panel) and mean charge state $\overline{q}$ (Equation~\ref{eq:qmean}, lower panel) resulting from photoionization of La$^+$.}
\end{figure}

Another discernible feature in the measured cross section is the threshold for $3p_{3/2}$ ionization at $\sim$1120~eV. It is most clearly visible in $\sigma_8$  and $\sigma_9$. The onset of $\sigma_{11}$ is most probably also related to $3p$ ionization. Apparently, the highest measured product-ion charge state $q=11$ cannot be reached after an initial $3d$ excitation or ionization. An increase of the cross sections $\sigma_q$ above the $3p_{3/2}$ threshold occurs only for $q\geq 7$. For lower $q$, particularly visible for $q=6$, the cross sections decrease above the  $3p_{3/2}$ threshold, i.e., creating a $3p$ hole leads to a higher mean product-ion charge 
\begin{equation}\label{eq:qmean}
\overline{q} = \frac{1}{\sigma_\Sigma}\sum_{q=4}^{12}q\sigma_q = \sum_{q=4}^{12}qf_q
\end{equation}
than the creation of an initial $3d$ hole as can directly be seen from the lower panel of Figure~\ref{fig:frac}.

The upper panel of Figure~\ref{fig:frac} displays the individual product-ion charge state fractions
\begin{equation}\label{eq:fq}
f_q = \frac{\sigma_q}{\sigma_\Sigma},
\end{equation}
none of which exceeds 50\%. In the energy region of the $3d\to4f$ resonances the dominant product-ion charge state is $q=5$, while at higher energies it is $q=6$. There, the lowest measured charge state, $q=3$, contributes by less than 5\%. La$^{2+}$ photoions could not be detected in the experimental energy range since the respective count rates were too small. This is due to the fact that the creation of an inner-shell hole is almost exclusively followed by a fast autoionizing transition. Beyond the $3p$ ionization threshold, charge states $q=5$, $q=6$, and $q=7$ occur with almost the same percentage (about 25\%), and also $q=8$ and $q=9$ become significant. The much smaller fractions of the higher measured charge states $q=10$ and $q=11$ are not shown in Figure~\ref{fig:frac}.

\begin{figure}
		\centering\includegraphics[width=\columnwidth]{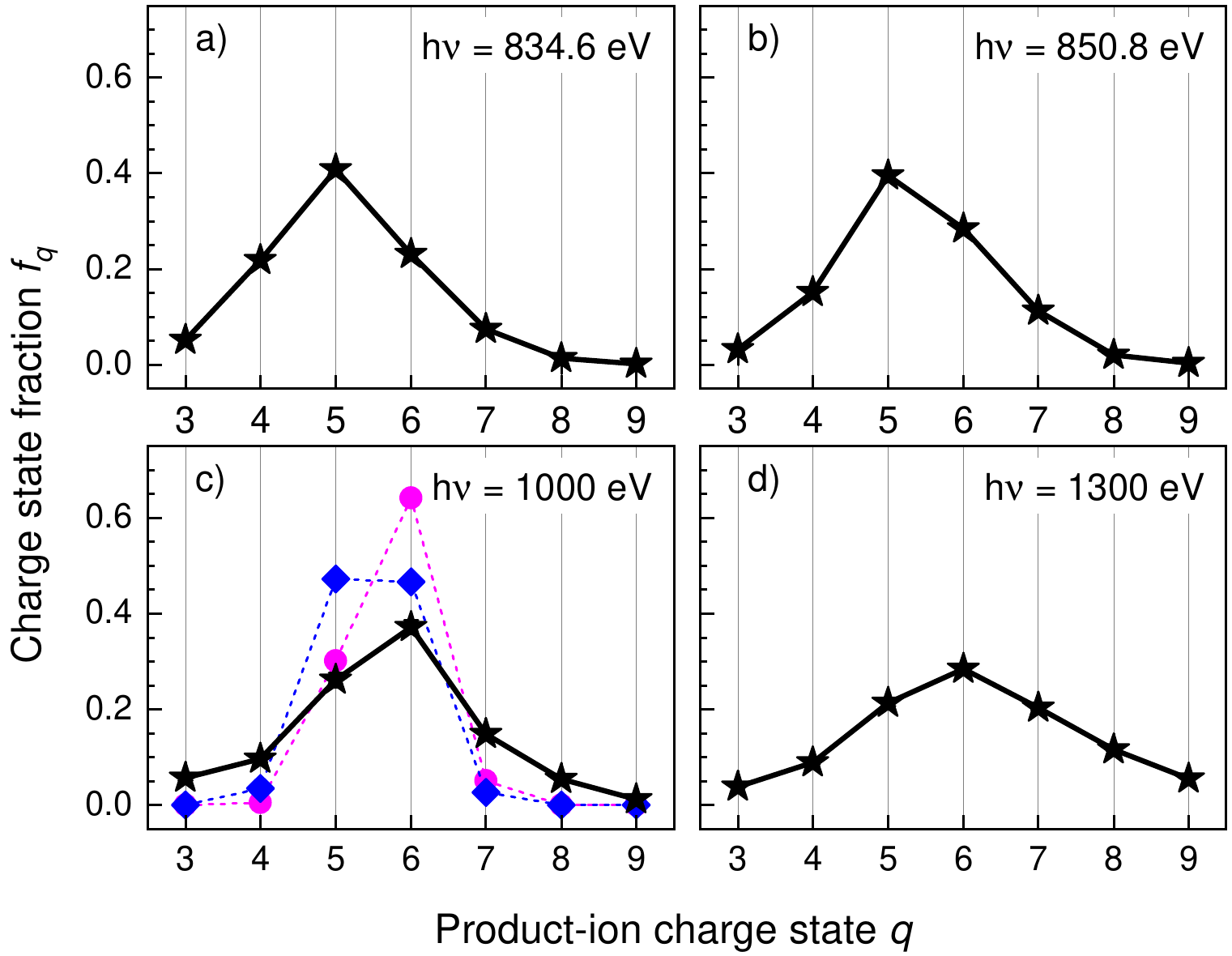}
	\caption{\label{fig:casc}Experimental (black stars connected by full lines) and theoretical (symbols connected by dashed lines) product\-ion charge state fractions $f_q$ (Equation~\ref{eq:fq}) for the photon energies $h\nu$ given. The experimental error bars are smaller than the size of the symbols. The computational models A (blue diamonds) and B (magenta circles) in panel c) are explained in the text. They predict mean charge states of 5.48 and 5.73, respectively.}
\end{figure}

Figure~\ref{fig:casc} visualizes selected charge state fractions from Figure~\ref{fig:frac} as functions of product-ion charge state. Panels a) and b) show the fractions at the positions of the two prominent resonance features in Figure~\ref{fig:all} that are associated with $3d_{5/2}$ and $3d_{3/2}$ excitations, respectively.  Panel c) shows experimental and theoretical (see below) charge state fractions at an energy above the $3d_{3/2}$ and below the $3p_{3/2}$ thresholds. The charge state fractions in panel d) were obtained at an energy well above the $3p_{3/2}$ threshold.  When going from lower to higher energies the charge distributions become flatter. This general trend has been observed already earlier \citep{Carlson1966a}. The charge state fractions at the two resonance energies peak at $q=5$ (panels a and b), whereas those beyond the $3d$ thresholds peak at $q=6$ (panels c and d). This is due to the fact that $3d$ ionization produces already a higher charge state in the primary process, whereas the charge state is not increased by an initial $3d$ excitation.    

To accurately predict the various cross sections for multiple ionization is even more demanding for theory than predicting the absorption cross section. This is because the calculation of the ionization cross sections additionally involves a detailed treatment of the deexcitation cascades which set in after the initial removal of an inner-shell electron \citep{Fritzsche2021,Fritzsche2024a}. Corresponding cascade calculations have already been carried out successfully with the JAC code for light ions such as O$^-$ \citep{Schippers2022a}, Si$^{1+}$, Si$^{2+}$, Si$^{3+}$ \citep{Schippers2022}, or Ar$^+$ \citep{Mueller2021b}. For the present work, we have extended the methodology to a much heavier atomic system with a significantly enlarged number of cascade steps as compared to the lighter systems just mentioned. 

In our present calculations, the cascades are initiated by the direct ionization of the $3d_{5/2}$ or $3d_{3/2}$ subshell, i.e., the calculations are valid in the photo-energy range of approximately 900–-1100~eV, above the $3d$ and below the $3p$ thresholds. We have considered the product charge states $2 \leq q \leq 10$, which involved a total of 1025 configurations and 141{,}245 cascade steps between these configurations. These cascade steps include 54{,}738 autoionization steps and 86{,}507 radiative emission steps. Because of the open-shell configurations of the selected ions, each step involves a larger number of transition lines between the pertaining fine-structure levels, thereby increasing the overall computational load. To maintain computational feasibility, we restrict our analysis to transitions with rates exceeding $1\times10^7$~s$^{-1}$. In addition, the JAC code employs parallel computing strategies \citep{Sahoo2024a} to substantially enhance the computational efficiency of these extensive calculations.

In Figure~\ref{fig:casc}c, we compare  our experimental charge state fractions (Figure~\ref{fig:frac}) for a photon energy of 1000~eV with results from two slightly different cascade models. Model A (blue diamonds) uses level energies as calculated by the JAC code. In Model B (magenta circles), the ionization energies of the selected charge states were adjusted to the compiled values reported in the NIST Atomic Spectra Data Base \citep{Kramida2024}, where applicable. Both models agree reasonably well with the experimental findings, however, the theoretical charge state distributions are somewhat narrower than the experimental one.  The mean product-ion charge states resulting from model~A and from model~B are 5.48 and 5.73, respectively. Within the experimental error bar, the value from model B agrees with the experimental mean charge state at 1000~eV of $5.66\pm0.14$. The mean charge state from model A is slightly too low.

The finding that the calculated charge-state distributions are narrower than the  experimental one (Figure~\ref{fig:casc}) is probably due to the present single-configuration approach, which neglects correlation effects, and due to the neglect of double Auger and shake processes, which were found to be influential for lighter ions \citep{Schippers2017,Beerwerth2019,Schippers2022a,Dong2025}, not to speak of triple Auger decay as found in the photoionization of C$^+$ ions \citep{Mueller2018b}. Moreover, in all our calculations we have disregarded that the ion beam  in our experiments consisted of a mixture of ground levels and metastable excited levels of the La$^+$ ion. Improving on some or even all of the deficiencies of the present cascade models would be computationally extremely demanding or even impossible with the present computing resources at hand.

\section{Summary and conclusions}\label{sec:sum}

The present experimental cross sections for multiple photoionization of La$^+$ ions in the energy range from below the $3d$ to above the $3p$ ionization thresholds are meant to serve as benchmarks for the quantum-theoretical calculations, which are used for generating the atomic data required for the nonequilibrium modeling of kilonovae. Our own theoretical calculations show that state-of-the-art atomic theory can reliably predict photoabsorption cross sections also for heavy many-electron atomic systems. Further code development is required for a more consistent treatment of nonresonant and resonant ionization.

The creation of a $3d$ hole by photoexcitation or direct photoionization initiates a deexcitation cascade consisting of numerous Auger and radiative transitions and eventually resulting in a rather broad distribution of product charge states. The sensitivity of our setup allowed us to record individual multiple ionization cross sections up to tenfold ionization by a \textit{single} photon. The present work is complementary to recent studies at x-ray free-electron lasers \citep{Richter2009a,Rudek2012,Roerig2023}, which report similarly high degrees of ionization for \textit{multi-photon} ionization of xenon atoms.   

Our theoretical treatment of the deexcitation cascade requires unprecedentedly large calculations. It reproduces the experimental mean product-ion charge state, however, the theoretical charge-state distributions are somewhat narrower than the experimental ones. This is most probably due to the current limitations of our cascade models, which were required to keep the computations tractable. In any case, our results show that ionization of heavy elements by energetic radiation entails a rather complex reaction network, which will have to be included in some detail in realistic nonequilibrium plasma modeling of kilonovae.            

\begin{acknowledgments}
We acknowledge DESY (Hamburg, Germany), a member of the Helmholtz Association HGF, for the provision of experimental facilities. Parts of this research were carried out at PETRA\,III and we would like to thank Frank Scholz and Moritz Hoesch for assistance in using beamline P04.
This research has been funded in part by the German Federal Ministry for Research, Technology and Space (BMFTR) within the ErUM-Pro funding scheme under contracts 05K25GU2,05K25RG1, and 05K25SJ1. M.M. acknowledges support by the Deutsche Forschungsgemeinschaft (DFG, Project No.\ 510114039). F.T. acknowledges funding by the DFG Emmy Noether Programme (Project No.\ 509471550). S.-X.~W.\ acknowledges support by the State of Hesse within the Research Cluster ELEMENTS (Project ID 500/10.006).
\end{acknowledgments}

\begin{contribution}

All authors contributed equally.


\end{contribution}

%

\facility{\pipe\ at beamline P04 of \piii\ at DESY, HPC cluster DRACO at the University of Jena}

\software{JAC \citep{Fritzsche2019}, \url{https://www.github.com/OpenJAC/JAC.jl}}  



\end{document}